\documentstyle[12pt]{article}

\begin{document}

\author{G.V.~Vlasov \\
Moscow Aviation Institute and Landau Institute for Theoretical Physics, \\
2 Kosygin~Street~117334,~Moscow~Russia\thanks{
E-mail: vs@itp.ac.ru}}
\title{Rotating relativistic superfluid}
\maketitle

\begin{abstract}
Relativistic equation of state and velocity comparable with the speed of
light are included in consideration of a superfluid rotating in a
cylindrical container. Minimizing the free energy, we derive the equation of
motion. It admits an analytic solution, the solid-body rotation inside and
irrotational motion near the walls of the vessel, providing the vortex
quantum is not extremely high that is satisfied for real astrophysical
objects. The relativistic velocity of the vessel and the relativistic
equation of state results to the deviation of the angular velocity of the
solid-body motion inside and that of the vessel. The boundary between the
solid-body and irrotational motion is also shifted sufficiently leading to a
difference between the total angular momentum of superfluid and the normal
matter.
\end{abstract}

\sloppy

\section{Introduction}

The purpose of the present study is to investigate the dynamics of the
relativistic rotating superfluid. It will be a generalization of the
non-relativistic theory \cite{Khalatnikov89}. The medium is usually called
relativistic in two senses: if it has a relativistic equation of state or
when it flows at a relativistic velocity. Both conditions (especially the
first one) take place in neutron stars whose massive interior is composed of
a superfluid nuclear matter \cite{ST83}. A superfluid with quantum vortices
has a third measure of relativity; it corresponds to the vortices - whether
they are relativistic~\cite{CL95}. This possibility pertains better to the
string fluids rather than to the real neutron star matter, although some
small relativistic effects are also feasible here. We do not omit them
immediately from the beginning and do not discuss more than it is necessary
in the present paper with the aim to find possible applications to the
neutron stars.

The signature of the Minkowsky space with a metric tensor $diag(\left\{
-+++\right\} $ and (if not specified otherwise) the natural system of units (%
$\hbar =c=1$) are used in the paper.

\section{A relativistic superfluid in a rotating cylinder}

{\bf 1.} Let a superfluid is contained in a cylindrical vessel rotating
along the axis $x^3$ with a constant angular velocity $\omega $ which is
assumed to be large enough for existence of a vortex array in a superfluid,
while the distance between neighboring vortices is small in comparison with
the size of container $R_0$. The equilibrium condition is determined by a
free energy minimization\cite{Khalatnikov89}:
\begin{equation}
\delta F=\delta E-\omega ^i\,\delta L_i=0  \label{free}
\end{equation}
where
\begin{equation}
E=\int T_0^0\,{\rm d}^3{\rm R}\qquad \omega ^i=\left( 0,0,\omega \right)
\label{energy}
\end{equation}
is the energy, while
\begin{equation}
L_i=\varepsilon _{ijk}L^{jk}\qquad L_3=-L^{12}  \label{angm}
\end{equation}
and \cite{BI97}
\begin{equation}
L^{jk}=\int \left( x^jT^{k0}-x^kT^{j0}\right) \,\,d\phi RdR  \label{mtens}
\end{equation}
is the angular momentum and its tensor form, respectively. The superfluid
velocity $V\left( R\right) $ at distance $R$ from the axis of rotation
follows immediately from the constraint $\delta F=0$.

{\bf 2.} In general the liquid and the vortices are not separated \cite{CL95}
due to the Lagrangian dependence $\Lambda $ on the amplitude
\begin{equation}
h^2=h^\nu h_\nu  \label{h}
\end{equation}
of the helicity vector
\begin{equation}
h^\nu =\,\frac 12\varepsilon ^{\nu \alpha \beta \gamma }\mu _\alpha W_{\beta
\gamma }  \label{hel}
\end{equation}
which, as we see, is the cross product of chemical potential 1-form $\mu
_\alpha =\mu u_\alpha $ and vorticity 2-form $W_{\beta \gamma }$. If $%
\partial \Lambda /\partial h=0$, the liquid and the vortices can be
considered separately (as in the frames of the dilatonic model \cite{CL95},
which may be treated as an analogue of the two-constituent superfluid model
\cite{Israel81}). Namely, the particle number current $n^\nu $ is collinear
to the chemical potential flow: $n^\nu =nu^\nu =\Phi ^2\mu ^\nu $. This
situation occurs in the weak vorticity limit which takes place inside a
typical neutron star~\cite{ST83}. Thus the energy-momentum tensor of
superfluid {\rm will} be \cite{CL95}
\begin{equation}
T_\rho ^\nu =\left( \rho +P\right) u^\nu u_\rho -\frac \lambda WW^{\nu
\sigma }W_{\sigma \rho }-\Psi g_\rho ^\nu  \label{tem}
\end{equation}
where
\begin{equation}
\lambda =K\Phi ^2\qquad \Phi ^2=\frac n\mu  \label{lambda}
\end{equation}
while pressure $P$ of the superfluid constituent and its rest-mass density $%
\rho $ satisfy the relation $\mu n=\rho +P$ pertaining to a perfect fluid.
The second term $\frac \lambda W$ in the right side of (\ref{tem}) is
negligible in the weak vorticity limit, while the pressure function $\Psi
\rightarrow P$.

{\bf 3.} For a vortex line situated at the distance $R$ from the axis of the
vessel it is convenient to switch to the reference frame rotating with the
superfluid medium at the velocity $V\left( R\right) $. The metric in the
rotating frame has the form \cite{LPPT75,LL88}
\begin{equation}
ds^2=-\left( 1-\omega ^2R^2\right) dt^2-2\omega Rd\varphi
\,dt+dR^2+R^2d\varphi ^2+dz^2  \label{interval}
\end{equation}
On account of small $b$ (with respect to the size of the vessel) the metric
in the vicinity $O\left( R\right) $ of the vortex line can be regarded as
flat. Thereby, we can also introduce the local cylindrical coordinates $%
\left( t,r,\phi ,z\right) $ and calculate the invariant circulation integral
\cite{CL95}
\begin{equation}
\frac 1{2\pi }\int\limits_{\partial U}\mu _\nu \,dx^\nu =\kappa =\hbar n
\label{circ}
\end{equation}
which determines the local superfluid velocity $v_\phi \left( r\right) $
round the vortex
\begin{equation}
v_\phi \left( r\right) =\frac{\mu _\phi }m\,=\frac \kappa {mr}=\frac{nl_c}r
\label{vsf}
\end{equation}
where $m$ is the mass of the boson pair and $l_c=\hbar /\left( mc\right) $
is the Compton length. Note that the superfluid velocity (defined as a
gradient of the bose-condensate wave function phase \cite{KL82}) does not
coincide with the usual four-velocity
\begin{equation}
u_\nu =\frac{\mu _\nu }\mu \,=\frac m\mu v_\nu  \label{four-us}
\end{equation}

This solution (locally) satisfies the irrotationality condition \cite
{Israel81,KL82}
\begin{equation}
{\rm rot}\,\vec v_s=0\qquad \mu _0={\rm const}  \label{rot}
\end{equation}
(where
\begin{equation}
{\rm rot}\vec v\equiv \frac 1R\partial _R\left( vR\right)  \label{rot-cyl}
\end{equation}
in cylindrical coordinates) which takes no place in the global sense,
particularly on the axis of the vortex line.

Although (in the local reference frame pinned to the vortex) the averaging
over domain $O\left( R\right) $ yields zero momentum
\begin{equation}
\langle \mu _\phi \rangle =0  \label{zero}
\end{equation}
the presence of a vortex results in the global rotation of the superfluid: $%
V\left( R\right) \neq 0$. The world sheet $\Sigma $ of the vortex line plays
the role of support of the vorticity 2-form $W_{\nu \varrho }$ ($W_{\nu
\varrho }\neq 0$ on $\Sigma $). Taking into account the link~(\ref{vsf})
between the superfluid velocity~$v_\phi $ and the flow chemical potential~$%
\mu _\phi $ and calculating the integral (\ref{circ}) in the laboratory
reference frame, we have
\begin{equation}
\kappa =\frac 1{2\pi }\oint\limits_{\partial U}\mu _\nu \,dx^\nu =\frac
1{2\pi }\int\limits_UW_{\nu \varrho }\,d\sigma ^{\nu \varrho }=\frac m{2\pi
}|{\rm rot}\vec v|\,\pi b^2  \label{circ2}
\end{equation}
where $|{\rm rot}\vec v|\,\left( R\right) $ is the value averaged over
domain $U\left( R\right) $. Indeed, the global velocity field $\vec V\left(
R\right) $ does not satisfy the irrotationality condition. Formula (\ref
{circ2}) implies relation between the velocity field and the distance
between the vortices:
\begin{equation}
b^2=\frac{2\kappa }{m|{\rm rot}\vec v|}  \label{b}
\end{equation}
Equations (\ref{circ})-(\ref{rot}) also imply
\begin{equation}
\mu ^2=-\mu ^\varrho \mu _\varrho =\mu _{*}^2-\frac{\kappa ^2}{r^2}
\label{mu}
\end{equation}
where $\mu _{*}\left( R\right) =\sqrt{\mu ^0\mu _0}=\mu _0/\sqrt{%
g_{00}\left( R\right) }$ and $\mu $ is invariant and can be treated as the
rest-frame chemical potential. The four-velocity~(\ref{four-us}) constructed
from (\ref{mu}) will be
\begin{equation}
u_\nu =\frac 1{\sqrt{1-w^2}}\left( -1,0,w,0\right) \qquad w\left( r\right)
=\frac \kappa {\mu r}=\frac{r_c}r  \label{four}
\end{equation}
where the modified Compton length is
\begin{equation}
r_c=\frac \kappa \mu =l_c\frac m\mu n  \label{four2}
\end{equation}

{\bf 4.} Followed the non-relativistic procedure \cite{Khalatnikov89}, we
can take into account the possible non-uniform distribution of vortices if:
calculate the energy of a single vortex line and multiply it on the density
of the vortices

\begin{equation}
N\left( R\right) =\frac 1{\pi b^2}=\frac{m|{\rm rot}\vec v|}{2\pi \kappa }
\label{n}
\end{equation}
Substituting (\ref{four}) in (\ref{tem}) we find the energy
\begin{equation}
E\left( R\right) =\int T_0^0\,rdr=\int\limits_a^b\left( \frac{\rho \left(
r\right) }{1-w^2}+\frac{w^2P}{1-w^2}\right) rdr  \label{ven}
\end{equation}
of a single vortex calculated in the local reference frame (associated with
point $R$). Of course, due to the local velocity $w\left( r\right) \neq 0$
round the vortex (\ref{four}) the mass density $\rho \left( r\right) $
deviates from the proper {\it rest} mass density $\rho _s$. Taking into
account the evident relation $\rho \left( r\right) =\rho _s\sqrt{1-w^2}$
between them and Eq.~(\ref{ven}), we obtain
\begin{equation}
E\left( R\right) =\pi \left[ \rho _s\left( b\sqrt{b^2-r_c^2}-a\sqrt{a^2-r_c^2%
}\right) +\rho _sr_c^2\ln \left( \frac{b+\sqrt{b^2-r_c^2}}{a+\sqrt{a^2-r_c^2}%
}\right) +Pr_c^2\ln \left( \frac{b^2-r_c^2}{a^2-r_c^2}\right) \right]
\label{ven2}
\end{equation}
where $a$ is the inner cutoff radius determined from the kinetics concepts.
The typical values for neutron stars are \cite{ST83}: $b\sim 10^{-4}\,$cm,
while $a$ is, at least, several times greater than $r_c\sim 1\,$fm (we do
not consider the exotic cluster vortices\thinspace \cite{Sedrakian95} with $%
n\sim 10^{12}$). Thus, Eq. (\ref{ven2}) is simplified so that
\begin{equation}
E\left( R\right) \cong \pi \left[ \rho _sb^2+\rho _sr_c^2\ln \left( \frac{2b%
}{a+\sqrt{a^2-r_c^2}}\right) +Pr_c^2\ln \left( \frac{b^2}{a^2-r_c^2}\right)
\right]  \label{ven3}
\end{equation}
Multiplying (\ref{ven3}) by (\ref{n}), we find the superfluid energy density
\begin{equation}
\varepsilon \left( R\right) =\rho _s+\rho _s\frac{r_c^2}{b^2}\ln \left(
\frac{2b}{a+\sqrt{a^2-r_c^2}}\right) +P\frac{r_c^2}{b^2}\ln \left( \frac{b^2%
}{a^2-r_c^2}\right)  \label{den}
\end{equation}
at point $R$, or (more roughly)
\begin{equation}
\varepsilon \left( R\right) =\rho _s+\left( \rho _s+2P\right) \frac{r_c^2}{%
b^2}\ln \left( \frac ba\right)  \label{den2}
\end{equation}
It is the mean value (we may define it also as $\varepsilon _s\equiv \langle
T_0^0\rangle $) of the energy density in the vicinity $O\left( R\right) $,
for (\ref{den}) bears no dependence on the local coordinate $r$ and, hence,
does not reflect the fine structure of the vortex cell. The energy $%
\varepsilon $ measured in the laboratory reference frame will be given
merely by formula \cite{LPPT75,LL88}
\begin{equation}
\varepsilon =\frac{\varepsilon \left( R\right) }{\sqrt{1-V_R^2}}
\label{den3}
\end{equation}
where Eq.~(\ref{zero}) was taken into account and the global ordinary
velocity $V_R$ is resulted from the global superfluid velocity (do not mix
it with the local filed~$v_\phi $)
\begin{equation}
v=\frac{V_R}{\sqrt{1-V_R^2}}  \label{V}
\end{equation}
which coincides with $V_R$ in the non-relativistic limit.

The mass density $\rho _L$ in the laboratory frame is expressed through mass
density in the rotating frame as $\rho _R=\rho _L\sqrt{1-V^2}$, while $\rho
_R$ is determined as the averaging over domain $O\left( R\right) $:
\begin{equation}
\rho _R=\langle \rho \left( r\right) \rangle =\int\limits_a^b\frac{\rho _s}{%
\sqrt{1-V_R^2}}\,rdr\simeq \rho _s\left( 1+\frac{r_c^2}{b^2}\ln \frac
ba\right)  \label{rho}
\end{equation}
Indeed, this term, which has appeared in (\ref{den2}), includes contribution
from the vortex.

Thereby
\begin{equation}
\varepsilon _R=\frac{\rho _s}{\sqrt{1-V_R{}^2}}+\frac{\rho _s+2P}{\sqrt{%
1-V_R^2}}\frac{r_c^2}{b\left( R\right) ^2}\ln \frac{b\left( R\right) }a
\label{den4}
\end{equation}
while, according to (\ref{vsf}), (\ref{b}), and (\ref{four2}),
\begin{equation}
\frac{r_c^2}{b^2}=\frac \kappa 2|{\rm rot}\vec v|\frac m{\mu ^2}=\frac{r_c}2|%
{\rm rot}\vec v|\frac m\mu =\frac \chi {4\pi }|{\rm rot}\vec v|  \label{rc}
\end{equation}
where
\begin{equation}
\chi =2\pi \hbar n\frac{m^2}{\mu ^2}  \label{chi}
\end{equation}
reduces to the usual circulation quantum\thinspace \cite{Khalatnikov89} in
the non-relativistic limit. We shall bellow a more convenient expression
\begin{equation}
\alpha =\frac \chi {8\pi }=cr_c\,\frac m{4\mu }=cl_c\,\frac n4\frac{m^2}{\mu
^2}  \label{alpha}
\end{equation}
instead.

{\bf 5.} The total angular momentum density \cite{BI97}
\begin{equation}
\left( 0,0,L_3\right) \rightarrow L^{12}{\bf =}M^{12}+S^{12}  \label{amom}
\end{equation}
includes spin contribution of the vortices $S^{jk}$ and the orbital part $%
M^{jk}$ from the rotating fluid itself. The latter may be calculated (in the
laboratory frame) by formula (\ref{mtens}):
\begin{equation}
M^{12}=\int x^1T^{20}\,dV=\int \left( \rho _R+P\right) \frac{V_R\,R}{\sqrt{%
1-V_R^2}}\,{\rm d}^3R  \label{orb}
\end{equation}
while the spin contribution from a single vortex is determined by integral (%
\ref{circ}):
\begin{equation}
s^{12}=\frac 1{2\pi }\oint\limits_{\partial O}\mu _\nu \,dx^\nu =\kappa
\label{spin1}
\end{equation}
It is invariant with respect to transformation from the rotating to the
laboratory reference frame. This is derived immediately from the constraints
\cite{LPPT75,LL88}
\begin{equation}
s^\mu s_\mu =-s^2\qquad p^\mu s_\mu =0  \label{spin-mom}
\end{equation}
which are the same in any reference frame; here $p^\mu $ is the momentum.
Since in the rotating frame
\begin{equation}
s_\mu =\frac 12\varepsilon _{\mu \nu \rho }\,s^{\nu \rho }=\left(
s^0,0,0,s^z\right) \qquad p^\mu =\left( p^0,0,p^\phi ,0\right)  \label{spin2}
\end{equation}
and $p^0\neq 0$, we find, by means of (\ref{spin-mom}) and metric (\ref
{interval}), that $s^0=0$. Hence, $s^2=\left( s^z\right) ^2$ and, in the
laboratory frame
\begin{equation}
\breve s_\mu =\frac 12\varepsilon _{\mu \nu \rho }\,\breve s^{\nu \rho
}=\left( \breve s^0,0,0,\breve s^z\right) \qquad p^\mu =\left( \breve
p^0,0,\breve p^\phi ,0\right)  \label{spin3}
\end{equation}
The identity (\ref{spin-mom}) then implies that
\begin{equation}
\breve p^0\breve s_0+\breve p^\phi \breve s_\phi =\breve p^0\breve s_0=0
\label{spin4}
\end{equation}
because $\breve s_0=0$, according to (\ref{spin4}), and the metric is
defined as (\ref{interval}). Therefore, also $\breve s_0=0$ and, hence, $%
\breve s_z^2=s^2=s_z^2$.

Thus, we can define the spin density
\begin{equation}
\frac{\rho _R}m\,s^{12}  \label{spin5}
\end{equation}
measured in the rotating reference frame (related to the point $R$).

Hence,
\begin{equation}
S^{12}=\int \frac{\kappa \,\rho _R/m}{\sqrt{1-V_R^2}}\,{\rm d}^3R
\label{spin6}
\end{equation}
will be the total spin in the laboratory frame~\cite{LPPT75,LL88}. Combining
Eqs.~(\ref{amom}), (\ref{orb}),~(\ref{spin6}) and~(\ref{rho}),~(\ref{rc}),
we get the total angular momentum
\begin{equation}
\frac L{2\pi \rho _s}=\ \int \frac{R{\rm d}R}{\sqrt{1-V_R^2}}\left\{ \left[
\left( 1+\frac{r_c^2}{b^2}\ln \frac ba\right) +\frac P{\rho _s}\right]
V_R\,R\,-\left( 1+\frac{r_c^2}{b^2}\ln \frac ba\right) \frac \kappa m\right\}
\label{mom-total}
\end{equation}

{\bf 6}. The total free energy is constructed from (\ref{free}), (\ref{den4}%
), (\ref{orb}), (\ref{amom}), (\ref{spin3}) so
\begin{equation}
\frac F{2\pi \rho _s}=\int \left\{ \left( 1+\frac{\rho _s+2P}{\rho _s}\frac{%
r_c^2}{b^2}\ln \frac ba\right) -\left[ \left( 1+\frac{r_c^2}{b^2}\ln \frac
ba\right) +\frac P{\rho _s}\right] \frac{v_R}c\omega R\,-\left( 1+\frac{r_c^2%
}{b^2}\ln \frac ba\right) \frac{\kappa \omega }m\right\} \frac{R{\rm d}R}{%
\sqrt{1-V_R^2}}  \label{free1}
\end{equation}
that is
\begin{equation}
\frac F{2\pi \rho _s}=\int RdR\left\{ J\gamma +\left( v\omega R-\Xi \gamma
\right) \frac \chi {8\pi }|{\rm rot}\vec v|\ln \frac{\pi |{\rm rot}\vec v|a^2%
}\chi -\Gamma v\omega R\right\}  \label{free5}
\end{equation}
where, in the ligth of~(\ref{rc}),
\begin{equation}
\gamma ={}\frac 1{\sqrt{1-V_R^2}}=\sqrt{1+v^2}  \label{g-factor}
\end{equation}
the parameters
\begin{equation}
J=1-\frac{\kappa \omega }m\qquad \Xi =1+\frac{2P}{\rho _s}-\frac{\kappa
\omega }m  \label{brief}
\end{equation}
and the equation of state (EOS) index
\begin{equation}
\Gamma =1+\frac P{\rho _s}  \label{index}
\end{equation}
tend to a unit in the non-relativistic limit. Note that
\begin{equation}
\xi =\frac{\kappa \omega }m=\frac{\hbar \omega n}{mc^2}=\frac{l_c}{R_2}\frac{%
n\omega }{\omega _M}\qquad \omega <\omega _M=\frac c{R_2}  \label{xi}
\end{equation}
implies the ratio of the spin constituent of the rotation energy per
particle (one may call it as ''rotation quantum'') to its rest-mass energy.
It does not exceed the ratio of the Compton length to the radius of the star
and occurs to be extremely small, namely, it is evaluated as $10^{-21}$ even
at angular velocity $\omega =10^3$ which is rather high for neutron stars.
One may only mediate at application to a hydrodynamic model of rotating
nuclei emphasized e.g. in~\cite{nucl}.

\section{The equation of motion}

Varying expression~(\ref{free5}) over $\delta v$, as shown in Appendix in
details, we obtain the equation of motion
\begin{equation}
J\,\frac v{\sqrt{1+v^2}}-\omega R\Gamma +\alpha \left[ \Xi \sqrt{1+v^2}%
-v\omega R\right] \frac{\partial _R|{\rm rot}\vec v|}{|{\rm rot}\vec v|}%
+\alpha \,\Xi \frac v{\sqrt{1+v^2}}\left[ |{\rm rot}\vec v|-\frac vR\ln
\frac{ea^2}{b^2}\right] -\alpha \omega R|{\rm rot}\vec v|=0  \label{motion1}
\end{equation}
Or, including the speed of light explicitly
\begin{equation}
\omega R\Gamma -J\,\frac v{\sqrt{1+v^2/c^2}}=\alpha \left[ \Xi \sqrt{%
1+v^2/c^2}-\frac 1{c^2}v\omega R\right] \frac{\partial _R|{\rm rot}\vec v|}{|%
{\rm rot}\vec v|}+\alpha \frac 1{c^2}\Xi \frac v{\sqrt{1+v^2/c^2}}\left(
\partial _Rv-\frac vR\ln \frac{b^2}{a^2}\right) -\alpha \omega R|{\rm rot}%
\vec v|  \label{motion3-light}
\end{equation}
While its dimensionless form
\begin{equation}
\left( J\,\frac v{\sqrt{1+v^2}}-x\Gamma \right) W=\xi _{*}\left\{ \Xi \frac
v{\sqrt{1+v^2}}\left( \frac vx\ln \frac{a^2}{b^2}-W\right) W+xW^2+\left(
W-\Xi \sqrt{1+v^2}\right) \partial _xW\right\}  \label{motion4-dim}
\end{equation}
with
\begin{equation}
\begin{array}{c}
\xi _{*}=\xi \frac{m^2}{4\mu ^2}\qquad W=\frac 1x\partial _x\left( xv\right)
\sim |{\rm rot}\vec v|\qquad x=\frac{\omega R}c
\end{array}
\label{dim}
\end{equation}
is also convenient for further discussion.

When simultaneously $c\rightarrow 0$ and $\Gamma \rightarrow 1$, Eq.~(\ref
{motion3-light}) reduices to the well-known non-relativistic equation~\cite
{Khalatnikov89}
\begin{equation}
v-\omega R+\alpha \,\frac{\partial _R|{\rm rot}\vec v|}{|{\rm rot}\vec v|}=0
\label{helium}
\end{equation}
for the rotating superfluid helium.

While the non-relativistic equation~(\ref{helium}) was describing the
solid-body rotation $v=\omega R$ in the inner domain ($R<R_i$) and the
irrotational motion

\begin{equation}
{\rm rot}\vec v=0  \label{irrot}
\end{equation}
in the outer domain ($R_i<R<R_2$), the general equation~(\ref{motion1})-(\ref
{motion4-dim}) also determines an irrotational solution~(\ref{irrot}) but it
does admit any solid-body rotation in the strict sense. However, when the
right side of~(\ref{motion4-dim}) is {\rm small}, the left side is splitted
into a product of these two solutions which are {\rm connected} in the
transition region ($R\simeq R_i$) whose width $l$ is small with respect to $%
R_i$ and $R_2-R_i$.

The intermediate region (between the solid-body rotation at small $R$ and
the irrotational motion in the outer layers) may become relatively broad at
significant right side of~(\ref{motion4-dim}). On account of small $\xi _{*}$%
, this may occur at ultra-relativistic velocities $v\sim 1/\xi \gg 1$. Even
if we apply the present analysis to the cluster vortices with $n\sim 10^{12}$%
, the lowest value will be
\begin{equation}
v>10^7  \label{ultra}
\end{equation}
Although one could consider this situation in the frames of the hydrodynamic
nulear model, it does not appear when we study the real neutron stars.
Therefore, except for the main transition region, the solution is determined
by equation
\begin{equation}
\left\{ \frac v{\sqrt{1+v^2}}-\omega R\Gamma \right\} \,{\rm rot}\vec
v=O\left( \xi \right)  \label{motion5}
\end{equation}
which yields
\begin{equation}
\frac{v_{+}}{\sqrt{1+v_{+}^2}}=V_{+}=\Gamma \omega R\qquad \sqrt{1+v_{+}^2}%
=\frac 1{\sqrt{1-\left( \Gamma \omega R\right) ^2}}  \label{sol}
\end{equation}
or, the angular velocity
\begin{equation}
\Omega =\Gamma \omega  \label{Omega}
\end{equation}
of the solid-body rotation in the inner region. It differs from the
non-relativistic rotation by multiple $\Gamma =1+\frac P{\rho _s}$. So, for
an ultrarelativistic equation of state ($\rho =3P$) a superfluid rotates at
angular velocity $\Omega =\frac 43\omega $ rather than $\Omega =\omega $.
For a stiff EOS ($\rho =P$) the superfluid angular velocity exceeds twice
the angular velocity of the vessel. The conclusion applied to the neutron
stars is evident: the angular velocity of the superfluid core $\Omega $
differs sufficiently from the angular velocity of the crust $\omega $.

As for the irrotational solution
\begin{equation}
v_{-}=\frac QR  \label{irr}
\end{equation}
of~(\ref{motion5}) in the outer region, it must obey the boundary condition $%
v\left( R_2\right) =\omega R_2/\sqrt{1-\omega ^2R_2^2/c^2}=Q/R_2$ and,
hence, be defined ultimately
\begin{equation}
v_{-}=\frac{\omega R_2}{\sqrt{1-\omega ^2R_2^2/c^2}}\frac{R_2}R  \label{irr2}
\end{equation}
That is
\begin{equation}
v_{-}=\frac{x_0^2}{x\sqrt{1-x_0^2}}=\frac qx  \label{x0}
\end{equation}
in the dimensionless form, where
\begin{equation}
q\equiv \ \frac{x_0^2}{\sqrt{1-x_0^2}}\qquad x_0\equiv \frac{\omega R_2}c
\label{q}
\end{equation}
Indeed, in the light of formula~(\ref{sol}) only~$x_0\leq 1/\Gamma $ has
physical sense.

\section{The boundary between two solutions}

\subsection{The extremum}

In order to find the boundary $R_i$ between the two types of motion,
determined by formulae~(\ref{irr2}) and~(\ref{sol}), respectively, we verify
the free energy~(\ref{F}) over unknown $R_i$ which must minimize $F$. The
latter, on account of the thin intermediate region (whose size is determined
by usual expression~\cite{Khalatnikov89} multiplied by $\Gamma $),{\rm \ }%
can be splitted into a sum
\begin{equation}
F=\int\limits_{R_i}^{R_2}f_{-}\left( R\right)
RdR+\int\limits_{R_1}^{R_i}f_{+}\left( R\right) \,RdR  \label{F}
\end{equation}
of irrotational $f_{-}\left( R\right) =f\left[ v_{-}\left( R\right) \right] $
and solid-body contribution $f_{+}\left( R\right) =f\left[ v_{+}\left(
R\right) \right] $, which we present in convenient dimensionless form
\begin{equation}
\frac{F_{-}}{2\pi \rho _s}=\frac{c^2}{\omega ^2}\int\limits_{x_i}^qx{\rm d}%
x\,\left( \sqrt{1+\frac{q^2}{x^2}}-\Gamma q\right)  \label{F-}
\end{equation}
\begin{equation}
\frac{F_{+}}{2\pi \rho _s}=\frac{c^2}{\omega ^2}\int\limits_0^{x_i}xdx\left%
\{ \frac 1{\sqrt{1-\Gamma ^2x^2}}-\frac{\Gamma ^2x^2}{\sqrt{1-\Gamma ^2x^2}}%
+o\left( \alpha \right) \right\}  \label{F+}
\end{equation}
where quantum contribution~$o\left( \alpha \right) $ is, evidently,
proportional to the quantum number $n$ (the non-linear dependence appears at
ultrarelativistic rotation, when the quantum term is sufficient, that may
occur in nuclear hydrodynamics).

The condition of extremum
\begin{equation}
\frac{dF}{dR}|_{R=R_i}=-f_{+}\left( R_i\right) R_i+f_{-}\left( R_i\right)
R_i=0  \label{extrem}
\end{equation}
or
\begin{equation}
f_{+}\left( x_i\right) =f_{-}\left( x_i\right)  \label{extrem-x}
\end{equation}
allows to determine radius $R_i=x_ic/\omega $ without direct calculation of
the total free energy~(\ref{F}),~(\ref{free}).

\subsection{The non-relativistic velocities}

In principle, we may confine ourselves with the non-relativistic velocities
of rotation (while the ratio $P/\rho $ is sufficient), because $\frac{\Omega
R_2}c$ $\leq \frac 1{30}$ at typical values $\Omega \leq 10^3~{\rm s}^{-1}$
and $R_2\simeq 10~{\rm km}$ for the majority of neutron stars. Even for very
high $\Omega $, approaching to $10^4\,{\rm s}^{-1}~$\cite{Prakash94}, the
quantity $x^2=\frac 19$ is small.

At non-relativistic velocities the free energy~(\ref{F}) reduces to
\begin{equation}
\frac{F_{nr}}{2\pi \rho _s}=\int RdR\left[ \frac{v^2}2-\Gamma v\omega R+\Xi
\alpha |{\rm rot}\vec v|\ln \frac{b^2}{a^2}\right]  \label{free-nr}
\end{equation}
and its irrotational and solid-body contribution are specified immediately:
\begin{equation}
f_{-}\left( R\right) =f\left[ v_{-}\left( R\right) \right] =\frac{\omega
^2R_2^2}2\frac{R_2^2}R-\Gamma \omega ^2R_2^2R  \label{nr-}
\end{equation}
\begin{equation}
f_{+}\left( R\right) =f\left[ v_{+}\left( R\right) \right] =\frac{\Omega
^2R^3}2+\Xi \alpha R|{\rm rot}\vec v|\ln \frac{b^2}{a^2}-\Gamma \Omega
\omega R^3  \label{nr+}
\end{equation}
Substituting them in extremum condition~(\ref{extrem}) leads to
\begin{equation}
\omega ^2\frac{R_2^2}{R_i^2}-2\Gamma \omega ^2=-\frac{\Omega ^2R_i^2}{R_2^2}+%
\frac{2\Xi \alpha }{R_2^2}\cdot 2\Omega \ln \frac{b^2}{a^2}  \label{extrem3}
\end{equation}
and, after plain arithmetics, yields{\rm \ }
\begin{equation}
R_i=\frac{R_2}{\sqrt{\Gamma }}-\sqrt{\frac{2\Gamma -1}\Gamma \frac \alpha
\omega \ln \frac{b^2}{a^2}}  \label{Ri}
\end{equation}
The quantum term in the right side of~(\ref{Ri}) is at least several orders
less than~$R_2$. But for superfluid helium~\cite{Khalatnikov89}~($\Gamma
-1=\frac P\rho \sim 10^{-16}$) it defines a narrow band
\begin{equation}
R_2-R_i=\sqrt{\frac \alpha \omega \ln \frac{b^2}{a^2}}\sim 10^{-2}{\rm \,cm}
\label{Ri-nr}
\end{equation}
of irrotational motion near the walls of the container. For a relativistic
matter, whose pressure $P$ is compared with its energy density $\rho $, the
deviation of the radius
\begin{equation}
R_i\cong \frac{R_2}{\sqrt{\Gamma }}  \label{Ri2}
\end{equation}
from $R_2$ is significant{\sc . }For an ultrarelativistic matter $%
R_i=0.87R_2 $, and $R_i=0.71R_2$ for a stiff matter ($\Gamma =2$).
Therefore, a $35\div 65~\%$ portion of the total volume (which is
proportional to $R^2$ in {\rm cylindrical} symmetry) is free of vortices.

\subsection{The relativistic velocities}

Substituting~(\ref{F-}) and~(\ref{F+}) in~(\ref{extrem-x}) we obtain the
equation
\begin{equation}
\sqrt{1+\frac{q^2}{x_i^2}}-\Gamma q=\sqrt{1-\Gamma ^2x_i^2}+o\left( \alpha
\right)  \label{x-extr}
\end{equation}
for dimensionless $x_i=\omega R_i/c$. As we have mentioned above, the
quantum term~$o\left( \alpha \right) $ does not play significant role in
defining $R_i$ at the relativistic equation of state. Solving~(\ref{F+})
without~$o\left( \alpha \right) $, we find that
\begin{equation}
\Gamma x_i^2=q\left( \sqrt{1+\frac{\Gamma ^2q^2}4}-\frac{\Gamma q}2\right) =%
\frac{x_0^2}{2\left( 1-x_0^2\right) }\left( \sqrt{4-4x_0^2+\Gamma ^2x_0^4}%
-\Gamma x_0^2\right)  \label{Ri3}
\end{equation}
generalizes~(\ref{Ri2}) at relativistic velocities. The ratio$~x_i/x_0$
always decreases with the growth of $x_0$ implying that relativistic
rotation tends to diminish $R_i$. And the ratio

\begin{equation}
\frac{x_i^2}{x_0^2}=\frac 1{\Gamma ^2-1}\left( \sqrt{\Gamma ^2-\frac 34}%
-\frac 12\right) ~  \label{Ri4}
\end{equation}
corresponds to the $\Gamma x_0\rightarrow 1$. Note, however, that~$%
x_i\rightarrow x_0$ as soon as~$\Gamma \rightarrow 1$; meanwhile, the
smallest ratio~$x_i/x_0=0.66$ is achieved at $\Gamma =2$.

\section{The angular momentum}

\subsection{The general expression}

The total angular momentum~(\ref{mom-total}) can be presented as a sum of
classical and quantum terms
\begin{equation}
\frac L{2\pi \rho _s}=\int R{\rm d}R\,\Gamma vR+L\left( \alpha \right)
\label{mom-tot2}
\end{equation}
where the quantum contribution $L\left( \alpha \right) $ is negligible in
comparison with the first term, because the right side of equation~(\ref
{motion4-dim}) is small (the quantum term is sufficient only at
ultrarelativistic rotation). Therefore, omitting it and substituting the
velocity fields~(\ref{irr2}) and~(\ref{sol}) in~(\ref{mom-tot2}), we get the
dimensionless expression
\begin{equation}
\frac L{2\pi \rho _s}\frac{\omega ^2}{\Gamma c^2}\simeq
\int\limits_0^{x_i}x^2dx\frac{\Gamma x}{\sqrt{1-\Gamma ^2x^2}}%
+\int\limits_{x_i}^{x_0}x^2dx\frac qx  \label{mom-appr}
\end{equation}
easily integrated up to
\begin{equation}
\frac L{2\pi \rho _s}\frac{\omega ^2}{\Gamma c^2}=\frac{2-\sqrt{1-\Gamma
^2x_i^2}\left( 2+\Gamma ^2x_i^2\right) }{3\Gamma ^3}+\frac 12q\left(
x_0^2-x_i^2\right)  \label{mom-dim}
\end{equation}
where $x_0=\omega R_2/c$ and $x_i$ are determined by\thinspace (\ref{Ri3}).

\subsection{The non-relativistic velocities}

The non-relativistic angular momentum can be derived either from formula~(%
\ref{mom-dim}) or from~(\ref{mom-tot2}) Substituting there~(\ref{irr2}) and~(%
\ref{sol}), we have
\begin{equation}
\frac{L_s}{2\pi \rho }=\int\limits_0^{R_i}\Omega R\Gamma \omega RR{\rm d}%
R+\int\limits_{R_i}^{R_2}\frac{\omega R_2^2}R\Gamma \omega RR{\rm d}R
\label{ls0}
\end{equation}
that is
\begin{equation}
\frac{L_s}{2\pi \rho }=\frac 12\omega ^2R_2^4\,\left( \Gamma -\frac
12\right) \qquad  \label{ls}
\end{equation}
and always exceeds the non-relativistic value $\frac L{2\pi \rho }=\frac
14\omega ^2R_2^4$. Their ratio is $5/3$ for $\Gamma =4/3$ and $3$ for $%
\Gamma =2$. Formula~(\ref{ls}) may be also compared with the angular
momentum
\begin{equation}
\frac{L_n}{2\pi \rho }=\frac 14\omega ^2R_2^4\,\Gamma ^2  \label{ln}
\end{equation}
of normal fluid rotating as a solid body with the same radius $R_2$ and
angular velocity $\omega $. If normal matter rotating with angular velocity $%
\omega _n$ transfers to superfluid with the same angular momentum, its
angular velocity
\begin{equation}
\frac{\omega _s^2}{\omega _n^2}=\frac{\Gamma ^2}{2\Gamma -1}  \label{solid}
\end{equation}
will be always greater than $\omega _n$. The first relativistic correction
\begin{equation}
\frac{\omega _s^2}{\omega _n^2}\cong \frac{\Gamma ^2}{2\Gamma -1}+\frac{%
\Gamma ^3}{2\Gamma -1}\left( \Gamma -\frac{4\Gamma -3}{\left( 2\Gamma
-1\right) ^{3/2}}\right) \,\frac{x_0^2}3+O\left( x_0^4\right)  \label{solid2}
\end{equation}
does not contradict to~(\ref{solid}) states the same, namely that the
superfluid rotates also {\it faster} than the normal fluid with the same
angular momentum. Meanwile,
\begin{equation}
\frac{\omega _s^2}{\omega _n^2}\rightarrow 1  \label{omega_s_n}
\end{equation}
when~$\Gamma \rightarrow 1$. Indeed, the difference between~$\omega _s$ and~$%
\omega _n$ is not negligible at relativistic EOS, because $\Gamma -1$ is
sufficient. This forms the background for further application to neutron
stars and pulsar glitches.

\subsection{The relativistic velocities}

The angular momentum of relativistic superfluid~(\ref{mom-dim}) differs from
its expression
\begin{equation}
\frac{L_n}{2\pi \rho _s}\frac{\omega ^2}{\Gamma c^2}=\frac{2-\sqrt{1-\Gamma
^2x_0^2}\,\left( 2+\Gamma ^2x_0^2\right) }{3\Gamma ^3}\qquad x_0\leq \frac
1\Gamma  \label{mom-norm}
\end{equation}
for normal matter with the same parameters, i.e. density~$\rho _s$, equation
of state (indicates $\Gamma $), external radius (dimensionless~$x_0$). The
angular momentum~(\ref{mom-norm}), as well as~(\ref{mom-dim}), inreases with
growth of$~x_0^2$. While always~$L_s\leq L_n$, especially at relativistic
rotation. Indeed, they coincide at~$\Gamma =1$ (for we have omitted the
negligible quntum contribution), while their ratio ($L_s/L_n$) at different~$%
\Gamma $ is given in the table below.
\begin{equation}
\begin{array}{cccccccccccc}
\begin{array}{c}
\Gamma x_0 \\
L_s/L_n
\end{array}
& 0 & 0.1 & 0.2 & 0.3 & 0.4 & 0.5 & 0.6 & 0.7 & 0.8 & 0.9 & 1 \\
\Gamma =1.1 & 0.992 & 0.99 & 0.989 & 0.987 & 0.984 & 0.979 & 0.973 & 0.964 &
0.947 & 0.911 & 0.694 \\
\Gamma =1.2 & 0.972 & 0.968 & 0.964 & 0.958 & 0.95 & 0.94 & 0.926 & 0.907 &
0.876 & 0.82 & 0.60 \\
\Gamma =1.6 & 0.86 & 0.85 & 0.83 & 0.82 & 0.80 & 0.78 & 0.75 & 0.71 & 0.67 &
0.60 & 0.42 \\
\,\Gamma =2 & 0.75 & 0.735 & 0.72 & 0.70 & 0.68 & 0.65 & 0.62 & 0.59 & 0.54
& 0.48 & 0.33
\end{array}
\label{table}
\end{equation}
Instead of~$x_0$, the universal variable~$\Gamma x_0$ stands here in the
first row.

\section{Conclusion}

Summarizing, we list the main results. Having derived the equation of
motion\thinspace (\ref{motion1})-(\ref{motion4-dim}) for the rotating
relativistic superfluid, we find that it may be reduced to a product of two
independent solutions\thinspace (\ref{motion5}), namely the solid body
rotation\thinspace (\ref{sol}) in the inner domain\thinspace $\,R<R_i$ and
the irrotational motion\thinspace (\ref{irr2}) in the outer domain\thinspace
\thinspace $R>R_i$ with a relatively thin region where they interfere at $%
R=R_i$. The splitting\thinspace (\ref{motion5}) into these independent
solutions\thinspace was possible due to a small rotation quantum\thinspace (%
\ref{xi}) implying that the velocity of rotation is not extremely
high\thinspace (\ref{ultra}), that takes place for all real macroscopic
objects, particularly, the neutron stars. Although, considering the nuclei
within the hydrodynamic approach, one encounters the opposite situation and
has to solve equation of motion\thinspace (\ref{motion4-dim}) numerically.

As for solutions\thinspace (\ref{sol}) and\thinspace (\ref{irr2}), there
should be noted the fact of dependence on the equation of state, besides the
relativistic expression for velocity. The {\it relativistic equation of state%
} and {\it relativistic rotation} are characterized by dimensionless
quantities~$\Gamma =1+P/\rho _s$ and~$x_0=\omega R_2/c$, respectively
(which, in the non-relativistic limit, become~$\Gamma \rightarrow 1$ and~$%
x_0\rightarrow 0$). However, without regard of rotation rate the
relativistic EOS results to difference between the angular
velocity\thinspace (\ref{Omega}) of the superfluid~-- obeying the solid-body
rotation\thinspace (\ref{sol}) within the vessel,~-- and the angular
velocity\thinspace $\omega $ of the vessel itself. In other words, the
superfluid rotating as a solid body rotates at angular velocity which$%
~\Gamma $ times higher than that of the vessel.

The same quantity~$\Gamma $ appears in formula~(\ref{Ri}),~(\ref{x-extr})~
for radius~$R_i$ which corresponds to the boundary between the solid-body
and irrotational motion. While the non-relativistic value of~$R_i$ deviates
slightly from~$R_2$ due to the presence of quantum term~(\ref{Ri-nr}), this
quantum term plays an insufficient role when the difference~(\ref{Ri2}),~~(%
\ref{Ri3}) between~$R_i$ and~$R_2$ becomes considerable on account of the
relativistic equation of state or relativistic rotation. Particularly, the
boundary radius~$R_i$ is determined merely by~$\Gamma $ for a relativistic
matter rotating at low velocity~(\ref{Ri2}). Or, briefly, Eq.~(\ref{Ri3})
states that both relativistic rotation and EOS {\it shorten} the distance~$%
R_i$, i.e. the region of irrotation motion is much {\it wider} than that in
non-relativistic helium.

So, the quantum term involved in the relevant non-relativistic formulae
becomes insufficient for a relativistic superfluid. Thus, the total angular
momentum~(\ref{mom-tot2}) is given approximately by formula (\ref{mom-dim}).
The angular momentum of a slowly rotating relativistic superfluid~(\ref{ls})
is always~less than the angular momentum~(\ref{ln}) of the normal fluid
rotating with the same angular frequency. Hence, if the normal matter,
rotating with frequency~$\omega _n$,\ transfers into a superfluid state
(with angular momentum conserved), its angular velocity~$\omega _s$\ will be
increased~(\ref{solid}). The general dependence (\ref{mom-dim}) of~$\omega
_s/\omega _n$\ on the initial angular velocity~$x_0=\omega R_2/c$\ and the
EOS index~$\Gamma $\ is more complicated as illustrated in table (\ref{table}%
). However, the angular momentum of superfluid~(\ref{mom-dim}) occurs to be
always smaller than that of normal matter~(\ref{mom-norm}) and the
relativistic rotation contributes to this tendency.

Thus, the equation of state and velocity~$\omega R_2$ of a rotating
superfluid are of particular importance when they belong to a relativistic
range. A plenty of applications (the pulsar glitches, for instance) are
expected to be derived from here. And further development in the light of
nuclear hydrodynamics or solution with a solid core (that~is$~R_2>R>R_1>0$)
may be also proposed.

\section{Appendix: variation of the free energy}

Since the rotation quantum~(\ref{xi}) is small for usual macroscopic
objects, we can rewrite~(\ref{free5}) as

\begin{equation}
\frac F{\rho _s}=2\pi \int f\,RdR\qquad f=\gamma \,J+\alpha |{\rm rot}\vec
v|\ln \frac{a^2}{b^2}\left\{ v\omega R-\psi \gamma \right\} -\Gamma v\omega R
\label{free6}
\end{equation}
Firstly we specify variation
\begin{equation}
\delta G\left[ v\right] =\frac{\partial G}{\partial v}\delta v\equiv
G^{\prime }\,\delta v  \label{g}
\end{equation}
of an arbitrary function~$G\left( v\right) $. Hence, the simplest
expressions for variations
\begin{equation}
\delta B=\delta \left( \gamma \,J-\Gamma v\omega R\right) =\left( \frac{J\,v%
}{\sqrt{1+v^2}}-\Gamma \omega R\right) \delta v  \label{s-var}
\end{equation}
for
\begin{equation}
\gamma =\frac 1{\sqrt{1-V_R^2}}=\sqrt{1+v^2}  \label{gamma}
\end{equation}
The variation of
\begin{equation}
A=-\alpha |{\rm rot}\vec v|\,G\ln \frac \chi {\pi |{\rm rot}\vec
v|a^2}=\alpha \,\frac{\partial _R\left( vR\right) }R\,G\ln \frac{a^2}{b^2}
\label{alfa}
\end{equation}
is performed so:
\begin{equation}
\delta A=\alpha \int \left[ \partial _R\left( \delta vR\right) G\ln \frac{a^2%
}{b^2}+\partial _R\left( vR\right) \frac{\delta \left( 1/b^2\right) }{1/b^2}%
G+\partial _R\left( vR\right) \ln \frac{a^2}{b^2}\delta G\right] dR\equiv
\delta A_1+\delta A_2+\delta A_3  \label{a-var}
\end{equation}
where
\begin{equation}
\ln \frac{a^2}{b^2}=\frac{\partial _R\left( 1/b^2\right) }{1/b^2}=\frac{%
\partial _R|{\rm rot}\vec v|}{|{\rm rot}\vec v|}\qquad \frac{\delta \left(
1/b^2\right) }{1/b^2}=\frac{\delta |{\rm rot}\vec v|}{|{\rm rot}\vec v|}=%
\frac{\partial _R\left( \delta vR\right) }{\partial _R\left( vR\right) }
\label{b-rot}
\end{equation}
Hence,
\begin{equation}
\delta A_2=\int \partial _R\left( \delta vR\right) GdR\qquad \delta A_3=\int
\partial _R\left( vR\right) \ln \frac{a^2}{b^2}G^{\prime }\,\delta v{\rm d}R
\label{b-rot2}
\end{equation}
The first two terms~$\delta A_1+\delta A_2$ of~(\ref{a-var}) are simplified
as
\begin{equation}
\delta A_1+\delta A_2=\int \partial _R\left( \delta vR\right) \ln \frac{ea^2%
}{b^2}GdR=\delta vR\ln \frac{ea^2}{b^2}G|_{R_1}^{R_2}-\int \delta vR\partial
_R\left( \ln \frac{ea^2}{b^2}G\right) dR  \label{free2}
\end{equation}
The total variation of~(\ref{a-var}) will be
\begin{equation}
\delta A=\delta vR\ln \frac{ea^2}{b^2}G|_{R_1}^{R_2}+\int \delta v\left\{
\partial _R\left( vR\right) \ln \frac{a^2}{b^2}\,G^{\prime }\left[ v\right]
-R\partial _R\left( G\left[ v\right] \ln \frac{ea^2}{b^2}\right) \right\} dR
\label{free7-0}
\end{equation}
Taking into account that variation $\delta v$\ vanishes at the edges of the
vessel (i.e. at $R=R_1$\ and $R=R_2$), and adding~(\ref{s-var}) to the
latter formula, we find the total variation of free energy
\begin{equation}
\delta F=\int \delta v\left[ \frac{J\,v}{\sqrt{1+v^2}}-\Gamma \omega
R+G^{\prime }\left( v\ln \frac{a^2}{b^2}-R\partial _Rv\right) -GR\frac{%
\partial _R|{\rm rot}\vec v|}{|{\rm rot}\vec v|}\right] dR
\end{equation}
which for arbitrary~$\delta v$ and$~G\left[ v\right] =v\omega R-\psi \gamma
\left[ v\right] $ yields equation of motion~(\ref{motion1}). The relevant
non-relativistic version is immediately obtained if we put $G\equiv 1$.

\end{document}